\documentclass[a4paper,12pt,fleqn]{article}

\usepackage{latexsym}
\usepackage{graphicx}
\usepackage{subfig}
\usepackage{amssymb}
\usepackage{amsmath}
\usepackage{amsfonts}

\begin{document}
\title{\bf Kolmogorov stochasticity parameter as a measure of quantum chaos}
\author{ Shashi C. L. Srivastava$^1$\footnote{Email: shashi@vecc.gov.in; Tel.: +91 33 23184462; Fax: +91 33 23346871} and Sudhir R. Jain$^2$\footnote{Email: srjain@barc.gov.in; Tel.: +91 22 25593589; Fax: +91 22 25505151}\\
$~^1$ {\em RIB Group, Variable Energy Cyclotron Centre} \\
{\em Kolkata 700064, India} \\
$~^{2}${\em Nuclear Physics Division, Bhabha Atomic Research Centre}\\
{\em Mumbai 400085, India} 
}
\date{}
\maketitle

\begin{abstract}

We propose the Kolmogorov stochasticity parameter, $\lambda $ for energy level spectra to classify quantum systems with corresponding classical dynamics ranging from integrable to chaotic. We also study the probability distribution function (PDF) of $\lambda $. Remarkably, the PDF of all the integrable systems studied here is the same and is found to be completely different from the PDF of chaotic systems. We also note that $\lambda _n$ for $n$ energy levels scales as $\lambda _n \sim n^{-\alpha }$. Furthermore, with $\alpha $, the stochastic probability (calculated from PDF) is seen to jump by about an order of magnitude as the systems turn chaotic. 
\end{abstract}
\vskip 0.5 truecm


\newpage
\section{Introduction}
Chaos in classical mechanics occurs as a result of loss of correlation between neighbouring initial conditions, leading to a very complex trajectory in phase space \cite{ott}. The faster this correlation is lost, more chaotic the system is and this is quantified by the Lyapunov exponent. However, in quantum mechanics, uncertainty principle does not allow a description in terms of trajectories which makes the notion of chaos in quantum systems to lose its meaning in the sense as it is used in classical mechanics. Moreover, unlike chaotic classical systems where one cannot obtain analytic expressions for classical trajectories, one can write (albeit rarely) such expressions for eigenfunctions \cite{eckhardt,jgk,ma,jain} of some quantum systems which are classically chaotic. In this paper, we study the Kolmogorov stochasticity parameter for level sequences and several related aspects in attempt to capture order-chaos transition in quantum mechanics. 

We consider billiards which are dynamical systems consisting of a freely moving particle in an enclosure, reflecting specularly from the boundary in accordance with Snell law. These systems possess a wide range of dynamical behaviour depending upon the shape of the enclosure. For instance, circular billiard enjoys two functionally independent constants of the motion, viz. angular momentum and energy, phase space trajectory resides on the surface of a two-dimensional torus. Such systems exhibit regular behaviour and are termed integrable systems if the dynamics is smooth everywhere \cite{arnold1}. However, integrability is easily broken by breakdown of symmetry or by presence of singularities giving rise to diffraction effects \cite{djm}. A classical trajectory of an integrable system can be written in terms of Fourier modes built on periods corresponding to the canonical angle variables, this is not possible for a chaotic system. Quantum studies \cite{brack} are concentrated on understanding the fluctuations in the level density in terms of invariant structures (e.g., periodic orbits) in phase space. However, for the billiards, the average level density is completely determined by the geometrical features like volume, area, perimeter, etc. owing to a well-known result of Weyl \cite{brack}. Thus, one concentrates on the fluctuations in the level density characterized by level correlations, which are eventually studied semiclassically in terms of periodic orbits by employing trace formulae. Perhaps the most popular measure is the nearest-neighbour level-spacing statistics \cite{linda,haake}, $P(S)$ giving the frequency of occurrence of certain spacing, $S$ for a given spectrum of levels. For generic integrable systems, it was argued by Berry and Tabor that the spacing among nearest neighbours is distributed in a Poissonian manner \cite{Berry}.

Statistical study of fluctuations of level density has a mathematical counterpart in statistics of eigenvalues of random matrices \cite{mehta} constructed within certain well-defined symmetry constraints. If a physical system is invariant under rotation and time reversal invariance, then the spectral statistics could be compared to the corresponding random matrices that are real-symmetric. It is observed that the fluctuation properties of energy levels of quantized time-reversal invariant chaotic systems agree with that of orthogonal ensemble of random matrices exhibiting level repulsion as $P(S) \sim S$, for small S \cite{haake,bgs,mehta}. However, a linear level repulsion is observed for quantized polygonal billiards \cite{djm,gremaud,bgs99,rb}  also even when these are non-integrable and non-chaotic \cite{rb,efv,jl}. This necessitates quest for a parameter which not only varies monotonically with underlying classical description but can also discriminate between chaoticity and non-integrability, independent of random matrix theory and semiclassical theories, however not in disagreement with them. 

One of the most significant advantages of this work is in its applicability to {\it finite} sequences of energy levels. In many physical situations, particularly in nuclear physics, the energy levels available are finite in number, of the order less than a hundred. In such situations, it becomes inappropriate to apply random matrix theory. We hope that this work will make a discussion of order-chaos transition in nuclei possible even for realistic systems with some tens of energy levels. Arnold \cite{arnold} has employed the stochasticity parameter of Kolmogorov \cite{kolmogorov} to number sequences of length in tens successfully. This work is an adaptation of the work by Kolmogorov and Arnold to energy levels of quantum systems in particular, and to a sequence of modes supported by any physical system in general.  

The plan of the paper is as follows. In Section 2, we present the definition of Kolmogorov stochasticity parameter, $\lambda $. Based on semiclassical trace formulae, semiclassical expressions for the parameter are given for different systems and these are compared with the numerical calculations. Putting all these expressions together, we show that $\lambda $ scales with the length of the set of energy levels. In Section 3, we recapitulate Kolmogorov statistic briefly and present some of the applications of this for number sequences, presented recently by Arnold \cite{arnold}. Subsequently, we present probability distribution functions (PDF) of $\lambda $ for energy levels of different dynamical systems. We also present results for zeros of Riemann zeta functions, and, for Gaussian unitary random matrix ensemble. Finally, we conclude by showing that with these PDFs, we can visualize an order-chaos transition in spectral properties of dynamical systems.   

\section{Kolmogorov stochasticity parameter}

\subsection{Definition}

Consider a quantum system ${\mathcal S}_i$ which possesses energy levels corresponding to its bound states, $E_1 \leq E_2 \leq \ldots E_n$. Cumulative density of levels is defined as 
\begin{equation}\label{eq:1}
N_n(E) = \sum_{i=1}^{n} g_i\Theta (E - E_i)
\end{equation}
where $g_i$ denotes the degree of degeneracy of the level, $E_i$. The average cumulative density of levels can be obtained by various methods for quantum systems. To be specific, we take simple systems that are illustrative and very popular in studies in quantum chaos - the two-dimensional quantum billiards and quantized maps. Quantum billiard is a system consisting of a point particle of mass $m$ enclosed in a box of a certain shape, ${\mathcal D}$; the solutions of the Schr\"{o}dinger equation are assumed to exist and the eigenvalues $\{E_i\}$ are obtained. For such systems, the average cumulative density of levels, $N_0(E)$ is given by Weyl formula \cite{baltes}.  

In analogy with the definition given by Kolmogorov \cite{footnote1}, let us define stochasticity parameter,
\begin{equation}\label{eq:2}
\lambda _n = \sup_{E}\frac{|N_n(E) - N_0(E)|}{\sqrt{n}}. 
\end{equation}
In the limit of $n \to \infty $, the difference $N_n(E) - N_0(E)$ will be just the oscillating part of the density of levels for which we can employ trace formulae \cite{brack} whenever possible. For the eigenvalues of a Laplacian (sometimes related to number theory), there are trace formulae which can be employed \cite{terras}, one of the well-known being the Selberg trace formula. This further helps us to show that even finite sequence length of size 5000 reproduces the behaviour obtained from trace formulae. In the Appendix, we have collected the numerical calculations of $\lambda _n$ and compared them with the results based on trace formulae. As can be seen there, these results agree well with each other for various billiard systems and maps. 

We would like to recall that statistical investigations of $N_n(E) - N_0(E)$ have been carried out in the past, albeit with a different normalization, as a measure of quantum chaos in spectra \cite{aurich}. In these studies, the measure was taken as 
\begin{equation}\label{eq:aurich}
W(x) = \frac{N_n(E) - N_0(E)}{\sqrt{\Delta _{\infty }(E)}}
\end{equation}  
where $\Delta _{\infty }(E)$ is the limit of the spectral rigidity, $\Delta _3(L;E)$ as $L \to \infty $. It was shown that $W(x)$ possesses a limit distribution with zero mean and unit variance for bound, conservative and scaling systems. For classically  (integrable) chaotic systems, the distribution was expected to be (non-)Gaussian. These predictions were verified in a statistical analysis of eigenmodes of superconducting microwave billiards \cite{alt}. For a Bunimovich stadium, the three-dimensional Sinai billiard, and for Limacon billiards, a Gaussian distribution was found. In contrast, for a two-dimensional circular billiard, a non-Gaussian distribution was found.   

Practically, to calculate $\lambda_n$ the numerical procedure followed is sketched as follows: 
(i) sort the given sequence; (ii) find out how many eigenvalues are less than or equal to $E$, from this subtract the expected number of such eigenvalues (calculated from Weyl's formula) and divide by square root of this number; (iii) vary $E$ and then take maximum value of number calculated from the last step, this is $\lambda_n$ for fixed sequence of length $n$.

\subsection{Semiclassical expressions}

\begin{figure}
\centering
\subfloat[rectangular billiard]{
\label{fig:sp_rect} 
\includegraphics[width=0.40\linewidth]{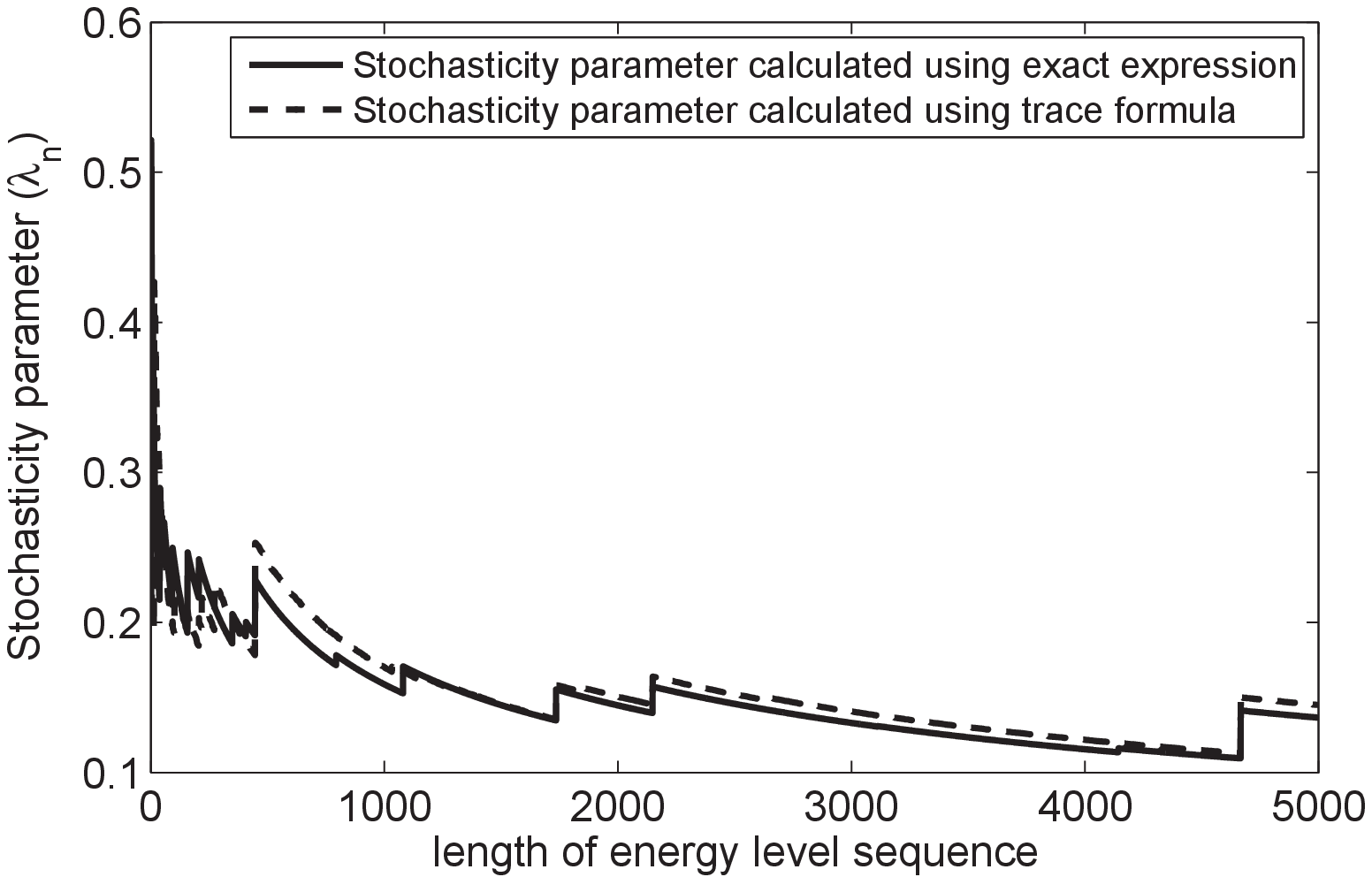}}
\hspace{0.05\linewidth}
\subfloat[circular billiard of unit radius]{
\label{fig:sp_circ} 
\includegraphics[width=0.40\linewidth]{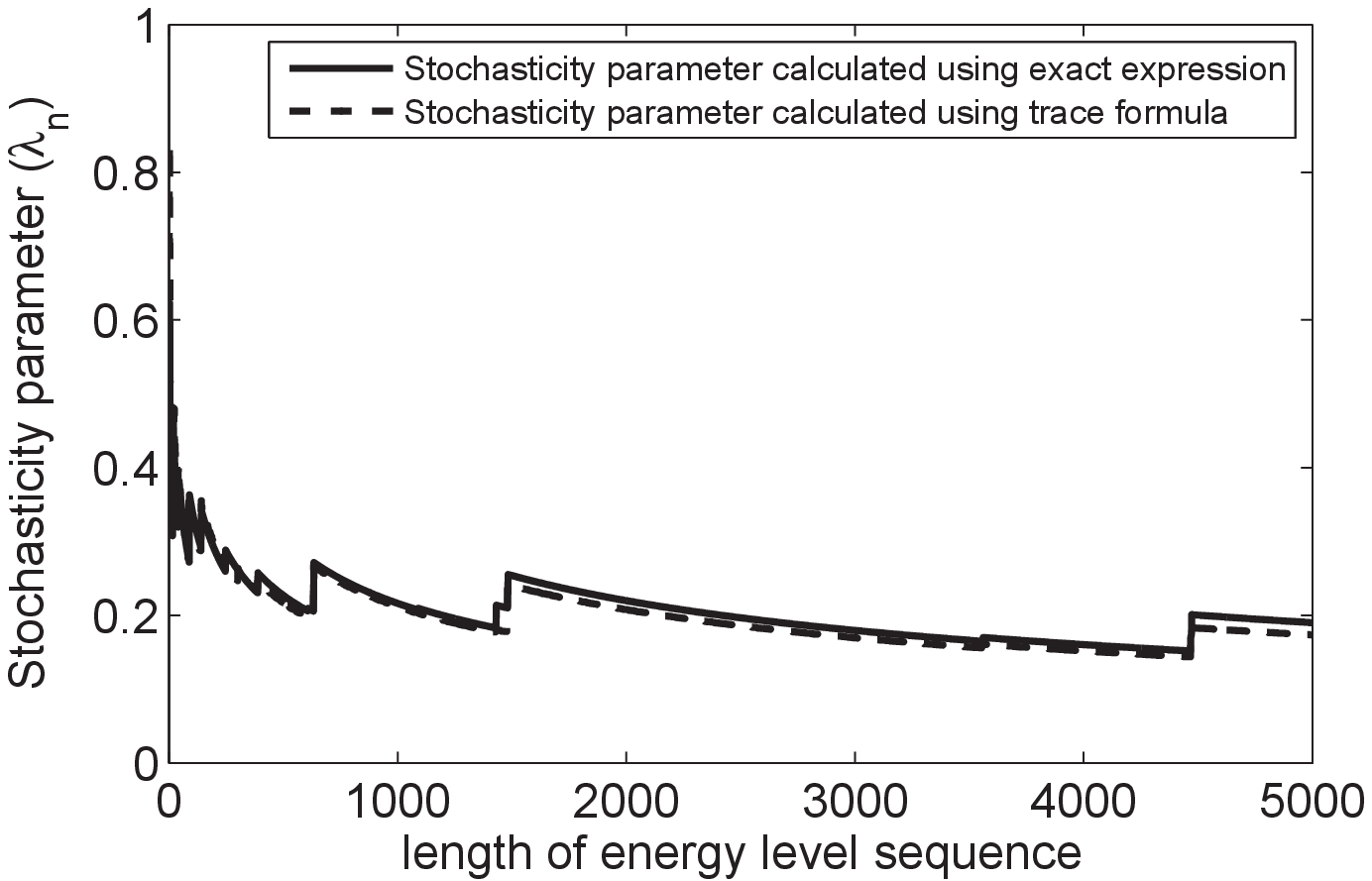}}\\[20pt]
\subfloat[triangular billiard of side length $\frac{4}{3}$]{
\label{fig:sp_tri} 
\includegraphics[width=0.40\linewidth]{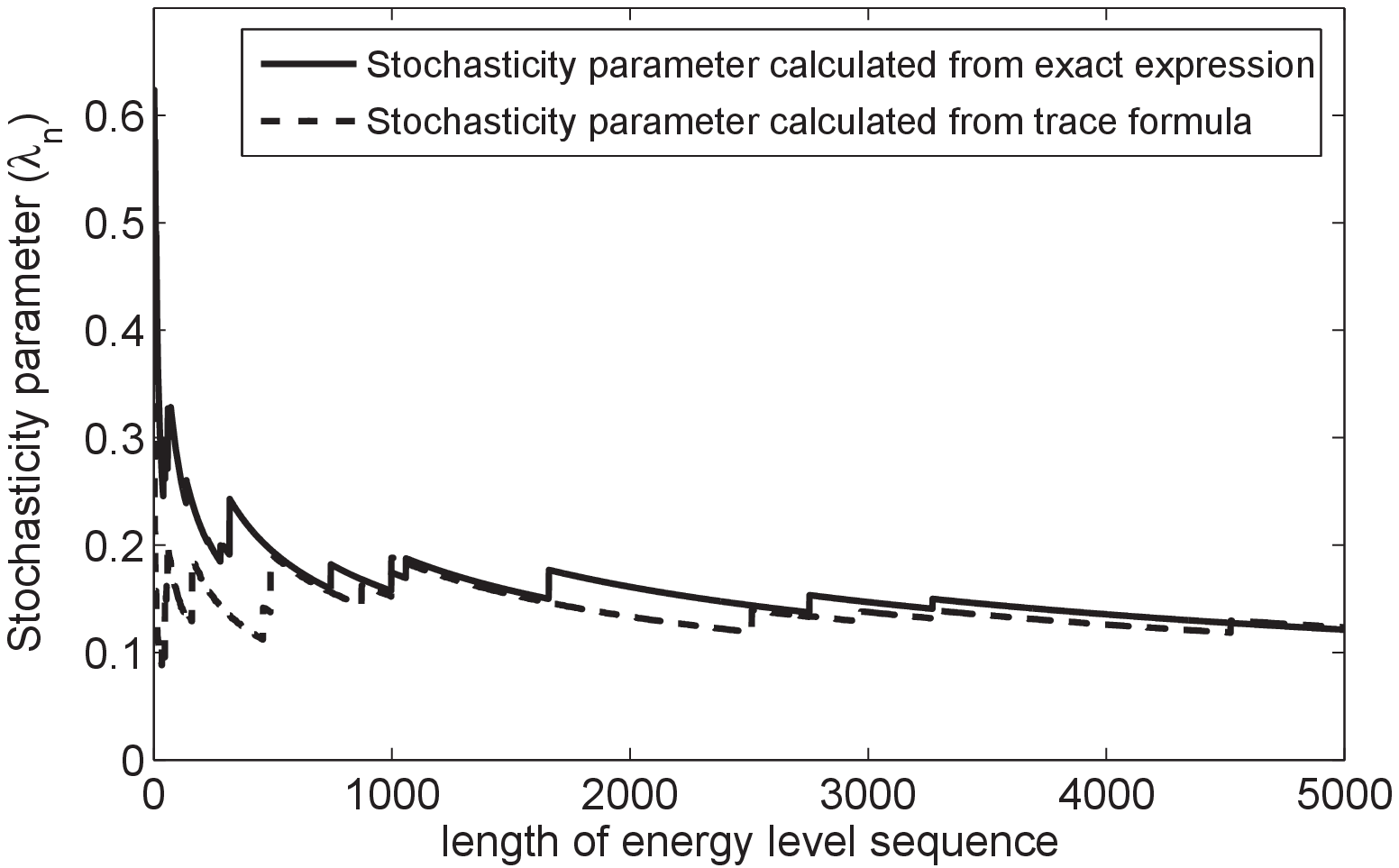}}
\hspace{0.05\linewidth}
\subfloat[zeros of Riemann zeta function,the values of $\sigma$ is taken as $\frac{1}{2}$]{
\label{fig:sp_zeta} 
\includegraphics[width=0.40\linewidth]{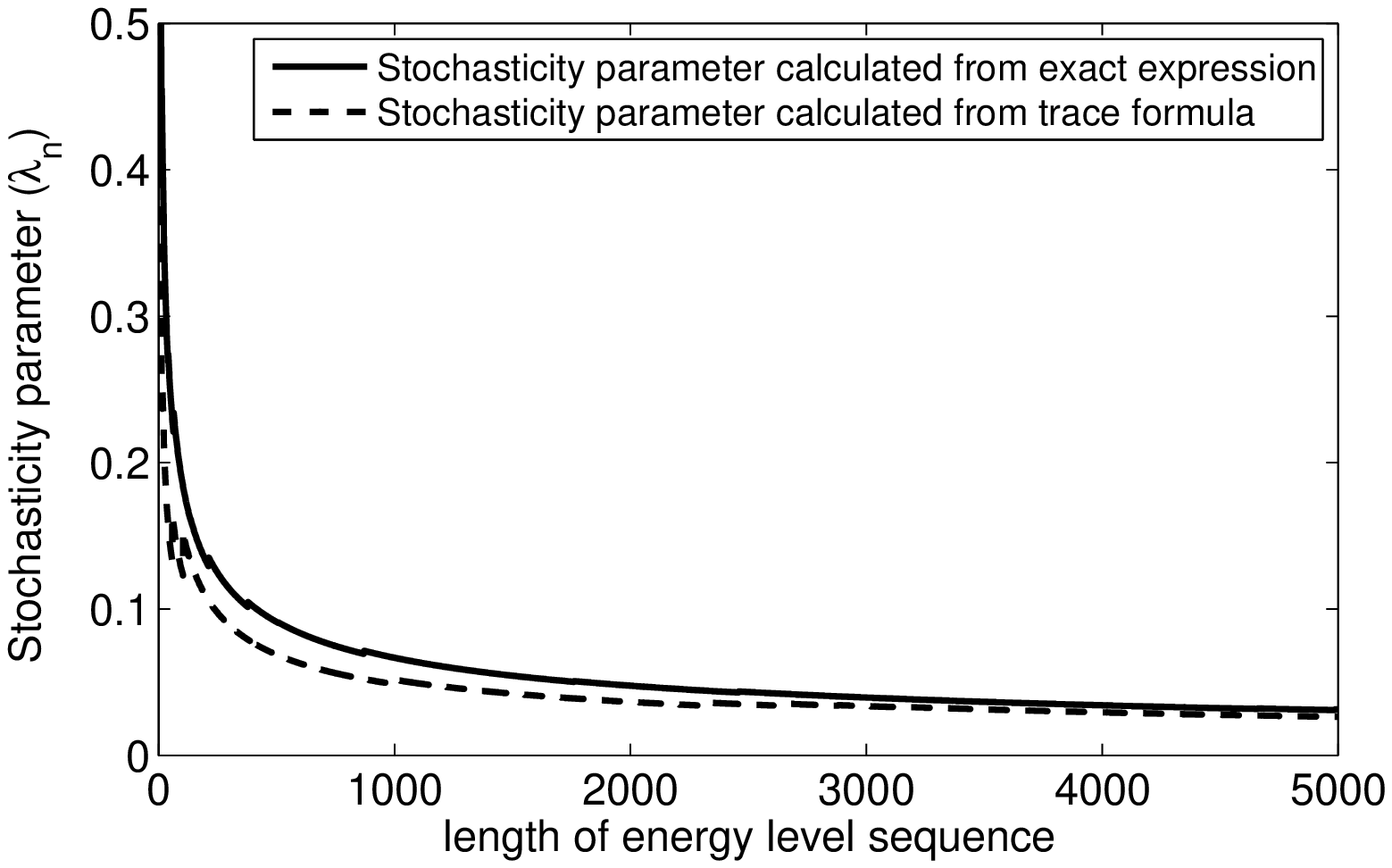}}
\caption[Stochasticity parameter]{Stochasticity parameter calculated using the exact expression and trace formula, the agreement is good.}
\label{fig:stopara} 
\end{figure}

Gutzwiller discovered the connection between density of energy levels of a quantum system and the periodic orbits of the underlying classical system \cite{Gutz1971}. The ensuing trace formulae have been found for quantum systems with underlying classical dynamics from regular to fully chaotic. For the harmonic oscillator potentials, the trace formula turns out to be analytically exact \cite{jain-brack}. In the following, we use the well-known trace formulae to get the Kolmogorov stochasticity parameter, $\lambda (E)$ and compare the expressions with numerical results. In the subsequent Sections, we will employ these to study probability density distribution functions of $\lambda (E)$. Thus, the agreement between the numerical and semiclassical results allow us to decide the length of the sequence required in studies in latter Sections. As seen in Fig. \ref{fig:stopara}, the agreement of numerical and semiclassical results is very good.

\subsection{Scaling}

\begin{figure}
\begin{center}
\includegraphics[scale=0.5]{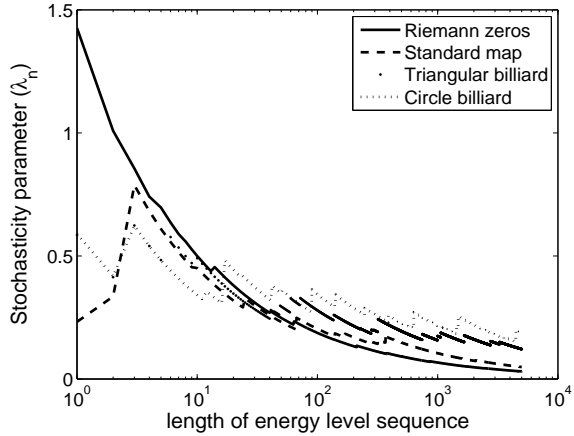}
\end{center}
\caption{Stochasticity parameter for all the systems considered are displayed here as a function of logarithm of the length of the sequence, $\ln n$. This brings out an interesting observation, $\lambda$ scales as $n^{-\alpha}$. This leads us to finding the best fitted values for the index, $\alpha$. As the systems become increasingly stochastic, the value of the index increases. }
\label{fig:def_alpha}
\end{figure}

Studying $\lambda_n$ for various systems discussed above as a function of $\ln n$ (Fig.\ref{fig:def_alpha}) reveals an exponential fall for Riemann zeros and weaker exponential curves for other systems. Indeed, on comparing each of them with $\exp(-\alpha \ln n)$, we extract a monotonically changing index, $\alpha$ varying from 0.15 for circular billiard to 0.45 for Riemann zeros with a value 0.34 for the chaotic standard map. The values of this index are listed in Table I. These show an increasing trend as the system becomes classically chaotic. 

We may understand the scaling exponent $\alpha $ heuristically in the following manner. 
For integrable systems, like rectangle the trace formula contains Bessel function and a  trigonometric function for $E$-dependence apart from $E^{-1/2}$ dependence in the denominator. For large $E$, $J_1(z) \sim \sqrt{2/ (\pi z) } \cos [z- 3 \pi /4]$. After a little simplification one will see that
\begin{eqnarray}
\lambda_n &\sim & \sup _E ~ |[ E^{-1/4}  (f_1 (\sqrt{E}) + f_2(\sqrt{E}) + f_3 
(\sqrt{E})] \nonumber \\& -& E^{-1/2} [ g_1 (\sqrt{E}) +g_2(\sqrt{E})]|.
\end{eqnarray}
where $f_i$'s and $g_i$'s are trigonometric functions ($\sin, \cos$) of $\sqrt{E}$. This leads to an scaling with a maximal exponent of about 0.25 in asymptotic limit with an oscillatory dependence with an argument containing $\sqrt{E}$. 

For the Riemann Zeta function it is just the denominator modified by the oscillatory behavior of oscillatory function which, to a good approximation may be written as  $t^{0.6}$.

Note that the above heuristic argument give a ratio of $\lambda $ for the zeta function (=0.6) to that for rectangle (=0.25) is 2.4 while the numerically found ratio 0.4451/0.18 is 2.47 - they are close. 

From \cite{aurich}, one may expect scaling behaviour because the systems considered here are scaling systems in the sense discussed above.

\section{Probability distribution functions }

\subsection{Kolmogorov statistic}

Since the average level density used in the definition of $\lambda$ is a continuous function, it renders a continuum character to $\lambda$ itself which in turn makes it possible to obtain the cumulative probability distribution function,$\Phi(\Lambda)$ of variables, $\lambda$. $\Phi(\Lambda)$ gives the probability of the stochasticity parameter to have a value $\lambda \leq \Lambda$. In the case of independent identically distributed random variables, this distribution is indeed a limiting distribution of Kolmogorov \cite{kendall},
\begin{equation}\label{kolmo}
\Phi(\Lambda) = \sum_ {k = -\infty}^\infty (-1)^k \exp (-2k^2 \Lambda ^2)
\end{equation}
It grows from $\Phi (0) = 0$ to $\Phi (+\infty ) = 1$. The mean value of $\Lambda $ is $\sqrt{\pi /2}\log 2 \sim 0.87$. This distribution has been used for comparing the degree of stochasticity in geometric and arithmetic progressions, among other examples \cite{arnold,arnold1}. Larger stochastic probability is obtained for a geometric progression than an arithmetic one, with only fifteen numbers. The stochastic probability of a sequence is quantified by the value $\Phi(\Lambda^0)$ where $\Lambda^0$ is obtained according to the definition above. 

\subsection{Extensions of Kolmogorov distribution for dynamical systems}

As suggested by Arnold \cite{arnold_ictp}, it will be very interesting to extend Kolmogorov formula to the more general situation for random variable having continuous cumulative distribution function. For the dynamical systems considered here, we have such a continuous distribution function. This gives us an opportunity to propose some of the extensions of Kolmogorov formula. We present results from our numerical investigations as well as analytic calculations based on trace formulae. 

The numerical evaluation of a probability distribution function is summarized in the following steps:
\begin{itemize}
\item for an ordered sequence of length $n$, we can have $n$-subsequences of length $1,2,\dots n$, always starting from the lowest element and increasing by one element;
\item for each subsequence we calculate $\lambda $;
\item this way, we end-up with $n$, $\lambda$s and a histogram of these $\lambda$s gives the probability distribution function (PDF) of $\lambda$ for the sequence.
\end{itemize}

\subsubsection{Integrable billiards}

\begin{figure}
\begin{center}
\includegraphics[height=2.0in]{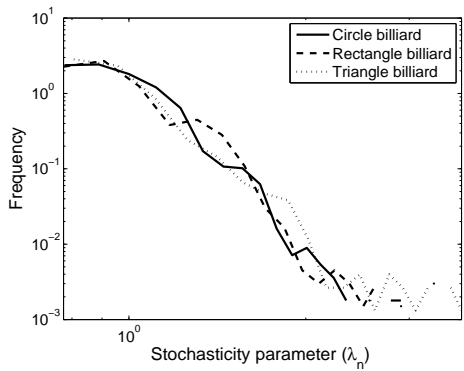}
\end{center}
\caption{The probability distribution of the stochasticity parameter is shown for three integrable billiards, viz. circle (red), rectangle (blue), and equilateral triangle (black). For these billiards, the distribution can be fitted to the same functional form, $c_1\exp(-\lambda^\beta))$, with ($c_1,\beta$) taking values (4.344, 2.927), (4.163, 3.196), and (4.703, 3.826) for circle, rectangle, and equilateral triangle respectively. Value of the  stochastic probability at $\Lambda$ measures quantum stochasticity. }
\label{fig:pdf_int}
\end{figure}

For the integrable billiards like circle, rectangle, and equilateral triangle, the values of the stochasticity parameter, $\lambda$ are respectively 0.19, 0.13, and 0.12, all comparable; these values are stable with the length of the sequence. Probability distribution function of $\lambda$ for the integrable systems are very close to each other (Fig. \ref{fig:pdf_int}) fitting the function, $\Phi'_{integrable}(\Lambda) \sim c_1 \exp(-\Lambda^\beta)$ (for details, see the Table. \ref{tab:sp}). The value of $\Phi$ at $\lambda$ gives the stochastic probability of the level sequence, in line with the measure of stochasticity for number sequences developed by Arnold \cite{arnold}. 
 
\begin{figure}
\begin{center}
\includegraphics[height=2.0in]{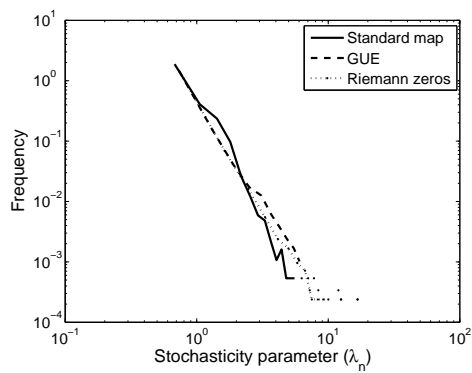}
\end{center}
\caption{The probability distribution of the stochasticity parameter is shown for standard map, GUE, and Riemann zeros. For these systems, the distribution can be fitted to the same functional form, $(c+\lambda)^{-\gamma}$, with ($c,\gamma$) taking values  (0.1688, 3.632), (0.1932, 4.097), and (0.2103, 4.397) for standard map, GUE, and Riemann zeros respectively. Value of the  stochastic probability at $\Lambda$ measures quantum stochasticity.}
\label{fig:pdf_chao}
\end{figure}

\subsubsection{Non-integrable systems}
 
For chaotic systems, as mentioned above, spectral fluctuations are found to be in close agreement with random matrix theory. A convincing argument for this statement has been given recently\cite{heusler} on the basis of Gutzwiller periodic orbit theory \cite{brack,haake}. For the quantized standard map, $\lambda_{standard}$ for 5000 levels is 0.0486. The function that fits the probability distribution of $\lambda$ for standard map, Riemann zeros, and eigenvalues of Gaussian Unitary Ensemble are all close to each other, represented by $(c+\lambda)^{-\gamma}$ where $\gamma$ is about 3.6 - 4.4  (see Table\ref{tab:sp}). We find the measure of stochasticity, $\Phi(\lambda_{standard}) = 0.4862$.  Further, on considering zeros of Riemann zeta function, $\lambda_\zeta = 0.0315$ and $\Phi(\lambda_\zeta) = 0.3775$. These are more than twice compared to the values for integrable case. Since it is very well-known that the fluctuation properties of Riemann zeros agree with those of the eigenvalues of unitary ensemble of random matrices, we have also calculated the stochasticity parameter and its distribution for them. We find $\lambda_{GUE} = 0.035$, fitting function is identical and $\Phi(\lambda_{GUE}) = 0.4028$. For the $\pi/3$-rhombus billiard, classical phase space surface is topologically equivalent to a sphere with two handles \cite{efv,jain}, Lyapunov exponent is zero. The energy levels of this almost integrable system entails an intermediate value for the stochasticity parameter. The probability distribution function also shows a concurrence with the family of integrable billiards (see Table \ref{tab:rhomb}).
	
\subsection{Order-chaos transition}

\begin{figure}
\begin{center}
\includegraphics[scale=0.5]{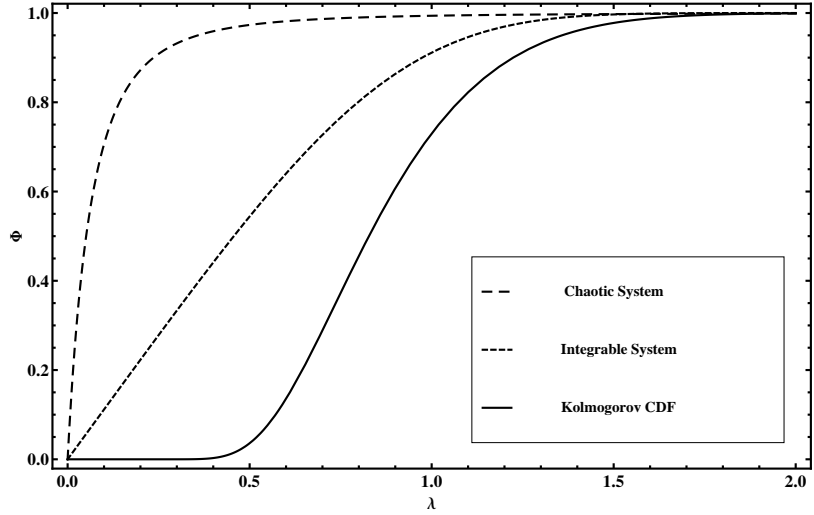}
\end{center}
\caption{A comparison among CDF for three different class of dynamical systems}
\label{fig:comp_cdf}
\end{figure}

\begin{figure}
\begin{center}
\includegraphics[scale=0.5]{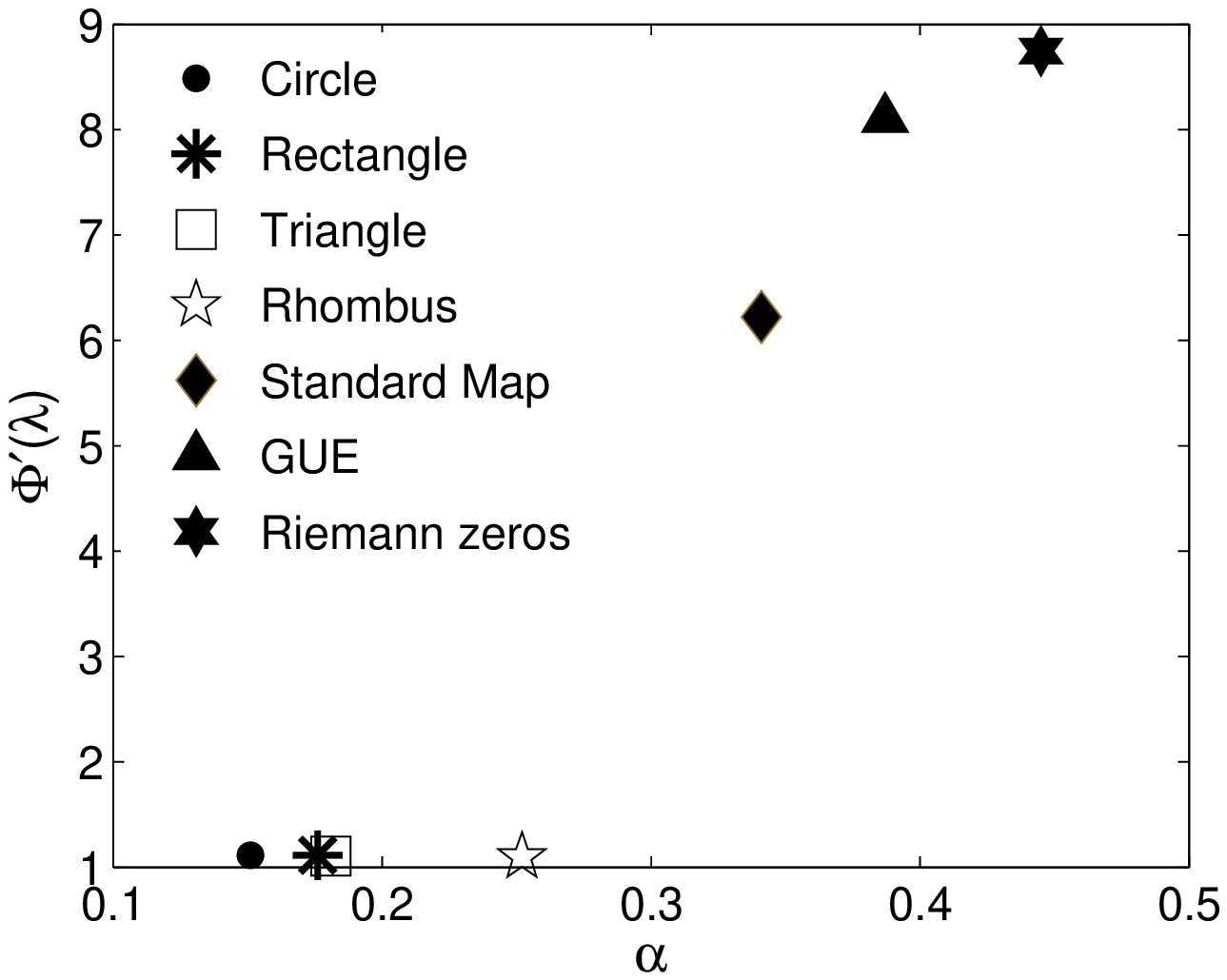}
\end{center}
\caption{Two measures of stochasticity are being plotted here against each other. For calculating the PDF at stochasticity parameter for a specific system, we have normalized the distributions. As the index $\alpha$ increases, we see that the probability distribution of the stochasticity parameter undergoes a jump by an order of magnitude. This $\Phi'-\alpha$ graph is a defining feature of quantum chaos.  }
\label{fig:phase}
\end{figure}

From the nature of probability distribution functions, we observe that the derivative of $\Phi$ at $\Lambda$ calculated for different systems jumps by six to eight times (Table \ref{tab:phase}), thus behaving like an �order parameter� for the order-chaos transition. The distinction among the cumulative distribution functions of integrable, chaotic and Kolmogorov distribution is apparent in Fig. \ref{fig:comp_cdf}. These results point at certain interesting conclusions: (i) stochasticity parameter can characterize the level sequences coming from random matrices, quantum systems, number theory etc., (ii) probability density distributions are distinctly different for integrable and chaotic quantum systems, (iii) we observe the confluence of the statistical behaviour of Riemann zeros and unitary ensemble in terms of $\lambda$ where we have not imposed any symmetry-related constraints as in random matrix theory. In a nutshell, we show (Fig.\ref{fig:phase}) the value of probability distribution function $\Phi'$ for all systems with $\alpha$. 

\begingroup
\begin{table}
\begin{center}
\caption[Phipalpha]{From the variation of stochasticity parameter, $\lambda$ with the length of sequence, $n$ (Fig. \ref{fig:def_alpha}), we extracted a scaling exponent $\alpha$. Employing the normalized cumulative distribution functions (CDF), we have also noted that the value of these are considerably larger for chaotic systems than for the integrable cases. As we go down the columns, we obtain larger values for the exponent and values of  $\Phi'$ increase too. However, it is the PDF (rightmost column) that jumps up to eight times which is enormous for the fact that all the distributions are normalized.}
\begin{tabular}{|p{2.0cm}| p{1.8cm}| p{1.8cm}|p{1.8cm}|}
\hline
System & $\alpha$ & $\Phi$ & $\Phi'$\\ \hline
Circle & 0.1509 & 0.2126 & 1.11239 \\ \hline
Rectangle &	0.1757	& 0.1523 &	1.11465 \\ \hline
Triangle & 0.1808 &	0.1340	 & 1.10569  \\ \hline
Rhombus	 & 0.2519	& 0.0479 &	1.09771 \\ \hline
Standard Map &	0.3409	 & 0.4862 &	6.22019 \\ \hline
GUE	& 0.3874 & 	0.4028 & 8.10375 \\ \hline
Riemann zeros &	0.4451	 & 0.3775 & 8.74426 \\ \hline

\end{tabular}

\label{tab:phase}
\end{center}
\end{table}
\endgroup

\begingroup
\begin{table}
\begin{center}
\caption[comparison]{Comparison of Stochasticity parameter for three different class of dynamical systems}
\begin{tabular}{|p{4.0cm}| p{4.0cm}| p{4.0cm}|}
\hline
\multicolumn{3}{|c|}{General model:$f(x) = a*\exp(-x^b)+0.0025$} \\
\hline
\multicolumn{1}{|c|}{Rectangle} &	Circle & Triangle \\ \hline
\multicolumn{1}{|p{4.0cm}|}{Coefficients (with 95 \% confidence bounds):
a=4.163 (3.575,4.75)
b=3.196 (2.202,4.19) 
Goodness of fit:
SSE: 2.33
R-square: 0.8721
 $\lambda_{5000}=0.1365$; $\Phi'(\lambda)=1.1146$} & Coefficients (with 95\% confidence bounds):
a=4.344 (3.907, 4.781)
b=2.927 (2.304, 3.551) 
Goodness of fit:
SSE: 1.509
R-square: 0.9227
$\lambda_{5000}=0.1900$; $\Phi'(\lambda)=1.1124$ & Coefficients (with 95\% confidence bounds):
a=4.703 (4.101, 5.305)
b=3.826 (2.719, 4.933) 
Goodness of fit:
SSE: 1.84
R-square: 0.9023
$\lambda_{5000}=0.1212$; $\Phi'(\lambda)=1.1057$ \\ \hline
\multicolumn{3}{|c|}{General model:$f(x) = (a+x)^{-b}$} \\
\hline
Standard Map &GUE & Riemann zeros\\ \hline
Coefficients (with 95\% confidence bounds):
a = 0.1688  (0.1606, 0.1771)  
b = 3.632  (3.463, 3.8)

Goodness of fit:
  SSE: 0.01436
  R-square: 0.9972
$\lambda_{5000}=0.0486$; $\Phi'(\lambda)=6.2202$
 &Coefficients (with 95 \% confidence bounds):
       a =    0.1932  (0.1907, 0.1958)    
       b =   4.097  (4.02, 4.173)  
Goodness of fit:
  SSE: 0.000886
  R-square: 0.9997
$\lambda_{5000}=0.035$; $\Phi'(\lambda)=8.1038$
 & Coefficients (with 95 \% confidence bounds):
       a = 0.2103 (0.2091, 0.2115) 
       b = 4.397 (4.336, 4.459) 
Goodness of fit:
  SSE: 0.0001541
  R-square: 0.9999
$\lambda_{5000}=0.0315$; $\Phi'(\lambda)=8.7443$ 
 \\ \hline
\end{tabular}

\label{tab:sp}
\end{center}
\end{table}

\endgroup

\begin{table}
\caption[fit_rhomb]{Fitting parameter for Rhombus}
\begin{center}
\begin{tabular}{|p{8.0cm}|}
\hline
General model:\\
       $~~~~~~~~f(x) = a*\exp(-x.^b)+0.0025$ \\
Coefficients (with 95 \% confidence bounds):\\ 
       \hspace{1cm}a =       3.626  (3.534, 3.718) \\
       \hspace{1cm}b =       4.367  (4.092, 4.642) \\
Goodness of fit: \\
  \hspace{1cm}SSE: 0.02527;  R-square: 0.9972;\\
   $~~~~~~~~\lambda_{5000}=0.0436$; $\Phi'(\lambda)=1.0977$\\
\hline
\end{tabular}
\label{tab:rhomb}
\end{center}
\end{table}

\subsection{Connections with number theory}

Finally, we would like to present the precise connection of our ideas with a well-known number-theoretic problem of counting the lattice points in a circle of radius, $R$, first studied by Gauss\cite{hardy}. The difference of this number with the area of the circle, known as the error term with proper normalization, has been investigated extensively\cite{iwan}. A connection of this problem with quantized integrable systems in two dimensions which have two quantum numbers has been studied in the past\cite{bleher}, however a measure of stochasticity was not discussed. The emergence of probability distribution functions of $\lambda $ for integrable and chaotic systems has been shown here. For cases that cannot be cast in terms of a problem with lattice points, our discussion opens up an enormous mathematical challenge to develop arguments which would probably result from a generalization of the classic work of Heath-Brown\cite{heath}.          

\section{Concluding remarks}

As the nature of classical behaviour of a system becomes more complicated, we have discovered the existence of a scaling exponent $\alpha$ of stochasticity parameter with the number of levels which follows the classical trend. Our studies are based on a 
number of systems displaying a variety of dynamical behaviour which have been shown to belong to distinct classes, not only in terms of this exponent but also in terms of the probability density distribution functions,  $\Phi'$ of the stochasticity parameter. The values of $\Phi'$ show a jump by a factor of six to eight as the system becomes chaotic. 

In his Kolmogorov lecture \cite{sinai}, Sinai explained that Renormalization Group method is connected to scaling ideas and limit theorems in probability theory. With the scaling property of $\lambda_n$ and existence of distinct PDFs, we may speculate a renormalization group argument to formally explain the transition noted in this work.   

\newpage 
\noindent
{\bf Appendix}

Here we collect the trace formulae \cite{brack} for different classes of dynamical systems and present the calculations of the stochsticity parameter based on them. The point is to make a comparison with the numerical results and we shall see that with 5000 energy levels, we have a good agreement (Fig. 1). We have chosen the following examples to guide us to the length of the sequence of levels required to arrive at reliable conclusions: (i) rectangle billiard - a separable, integrable system where the quantization condition is an algebraic relation; (ii) circular billiard - a separable, integrable system where the quantization condition is a transcendental equation; (iii) equilateral triangle billiard - a non-separable, integrable system where the quantization condition is an algebraic relation; (iv) Riemann zeta function - nontrivial Riemann zeros are being considered here, the interest being due to their possible connection with unitary ensemble of random matrices which, in turn, share spectral fluctuations with some well-known quantum systems which are classically chaotic.

\noindent
{\it Rectangle billiard}

Let us consider a rectangle billiard where the sides of the rectangle are $a_1$ and $a_2$ for which the density of energy levels (derivative of the cumulative density) is given by the trace formula \cite{brack}:
\begin{eqnarray}\label{eq:3}
d(E) &=& \sum_{i=1}^{\infty} g_i \delta (E - E_i) \nonumber \\
&=& \frac{ma_1a_2}{2\pi \hbar ^2}\sum_{(M_1,M_2)=(-\infty ,-\infty )}^{(\infty ,\infty )} J_0\left( \frac{S_{M_1M_2}}{\hbar } \right) \nonumber \\
&~& - \sum_{i=1 ,2 } \frac{a_i}{4 \pi \hbar}\sqrt{\frac{2 m}{E}}\sum_{M= -\infty}^{\infty} \cos\left( 2M a_i\sqrt{2mE}/\hbar\right).
\end{eqnarray}
where $J_0(x)$ is the cylindrical Bessel function \cite{edwards} and action, $S_{M_1M_2}$ is given by 
\begin{equation}\label{eq:4}
S_{M_1M_2} = \sqrt{2mE}L_{M_1M_2}, ~L_{M_1M_2} = \sqrt{M_1^2a_1^2 + M_2^2a_2^2}.
\end{equation}
With this, putting $\frac{\hbar^2}{2m} = 1$ (for all the systems),  
\begin{eqnarray}\label{eq:5}
N_n(E) - N_0(E) &=& \frac{a_1a_2}{4\pi } \sum_{(M_1,M_2)}^{(n_1,n_2)} \frac{8\sqrt{E}}{L_{M_1,M_2}}J_1[\sqrt{E}L_{M_1,M_2}] \nonumber \\
&~& +\frac{a_1a_2}{4\pi }\sum_{M_1=1}^\infty \frac{4\sqrt{E}}{L_{M_1,0}}J_1[\sqrt{E}L_{M_1,0}] \nonumber \\
&~&+\frac{a_1a_2}{4\pi }\sum_{M_2=1}^\infty \frac{4\sqrt{E}}{L_{0,M_2}}J_1[\sqrt{E}L_{0,M_2}] \nonumber \\
&~&-\frac{1}{4\pi }\sum_{M=1}^{\infty} \frac{2}{M}\left( \sin (2M a_1 \sqrt E )+\sin (2 M a_2 \sqrt E )\right)
\end{eqnarray}
where $J_1(x)$ is cylindrical Bessel function of order one.

For the denominator of (\ref{eq:2}), we will use Weyl's staircase function for the energy levels of rectangle. With the corner corrections included, this is read as
\begin{eqnarray}\label{eq:6}
n(E\leq E_0) &=& \frac{E_0}{4}-\frac{(1+\pi)\sqrt E_0}{2 \pi} + \frac{1}{4}.
\end{eqnarray}
where area and perimeter is taken $\pi$, $2(1+\pi)$ respectively.

Using first $5000$ energy levels of rectangle, we calculated Kolmogorov's stochasticity parameter (\ref{eq:2}) as a function of length of energy level sequence and compared the same when (\ref{eq:2}) is calculated using (\ref{eq:5}) and (\ref{eq:6}). The agreement is shown in Fig.\ref{fig:sp_rect}.

\noindent
{\it Circular billiard}

For a circular billiard of radius $R$, the treatment in terms of periodic orbits can be seen in the work of Balian and Bloch \cite{balian}. However, the trace formula was given by Bogachek and Gogadze \cite{bogachek}. The oscillatory part of the density of eigenvalues is 
\begin{equation}\label{eq:circle}
g_{osc}(E) = \frac{1}{E_o}\sqrt{\frac{\hbar }{\pi p R}} \sum_{w=1}^{\infty}\sum_{v=2w}^{\infty} f_{vw}\frac{\sin ^{\frac{3}{2}}\phi  _{vw}}{\sqrt{v}}\sin \Phi _{vw},
\end{equation}
where 
\begin{eqnarray}
E_o &=& \frac{\hbar ^2}{2mR^2}, ~~\phi _{vw} = \frac{\pi w}{v}, ~~L_{vw} = 2vR\sin \phi _{vw}, \nonumber \\
p &=& \sqrt{2mE}, ~~\Phi _{vw} = \frac{pL_{vw}}{\hbar } - 3v\frac{\pi }{2} + \frac{3\pi }{4}, \nonumber \\
f_{vw} &=&  1 \mbox{~~for~~} v = 2w \nonumber \\
&=& 2 \mbox{~~for~~} v > 2w.
\end{eqnarray}

The average cumulative density of energy, $E_0$ for a circle of area $\sigma $, circumference, $\gamma$ is given by \cite{baltes} 
\begin{equation}\label{eq:weyl_circle}
n(E\leq E_0) = \frac{\sigma E_0}{4\pi } - \frac{\gamma \sqrt{E_0}}{4\pi } + \frac{1}{6}. 
\end{equation}

\begin{eqnarray}
N_n(E) - N_0(E) &=& \frac{2mR^2}{\sqrt{\pi R}}\sum_{w=1}^{\infty} \int_{0}^{E} dE' \frac{\sin \left[ 4wR\sqrt{2mE'} - 3\pi w + \frac{3\pi }{4} \right]}{(2mE')^{\frac{1}{4}}} \nonumber \\ 
&~&+ \frac{4mR^2}{\sqrt{\pi R}}\sum_{w=1}^{\infty } \sum_{v > 2w} \frac{1}{\sqrt{v}}\sin ^{\frac{3}{2}}\left( \frac{\pi w}{v} \right) \nonumber \\
&~& 
\int_{0}^{E} dE' \frac{\sin \left[ 2vR \sin \frac{\pi w}{v}\sqrt{2mE'} - 3v\frac{\pi }{2} + \frac{3\pi }{4}\right]}{(2mE')^{\frac{1}{4}}}.
\end{eqnarray}
The integral involved in this expression is 
\begin{equation}
I(E) = \int_{0}^{E} dx x^{-\frac{1}{4}}\sin (\gamma \sqrt{x} + \delta ). 
\end{equation}
Let $\sqrt{x} = (y - \delta )/\gamma $, so $dx/(2\sqrt{x}) = dy/\gamma $. Then, $I(E)$ becomes
 
 \begin{eqnarray}\label{eq:Ncirc}
I(E) &=& \int_{\delta }^{\gamma \sqrt{E} + \delta } dy \frac{2}{\gamma^{3/2}}\sin y (y - \delta )^{1/2} \nonumber \\
&=& \int_{0}^{\gamma \sqrt{E}} \frac{2}{\gamma^{3/2}}\sin (z + \delta )z^{1/2} ~~ (z = y-\delta ) \nonumber \\
&=& -\frac{2 E^{1/4}}{\gamma} \cos(\gamma \sqrt{E}+\delta) + \sqrt{\frac{2\pi}{\gamma^3}}\left(\cos \delta \mbox{ FresnelC}\left[\sqrt{\frac{2\gamma \sqrt{E}}{\pi}}\right]\right. \nonumber \\
&~&~~~~~~~~~~~~~~~~\left. - \mbox{FresnelS}\left[\sqrt{\frac{2\gamma \sqrt{E}}{\pi}}\right]\sin \delta \right)
\end{eqnarray}
In (\ref{eq:Ncirc}), the terms containing FresnelC and FresnelS terms contributes at the third decimal place which for all the numerical comparison purposes we can ignore. The agreement is good and given in Fig.\ref{fig:sp_circ}.

\noindent
{\it Equilateral triangular billiard}

Equilateral triangular billiard represents a class of dynamical systems which are integrable but not separable. For equilateral triangular billiard, the eigenvalue spectrum is given by
\begin{eqnarray}\label{eq:trilevel}
E(m,n) &=& E_\Delta \left( m^2+n^2-mn \right) \nonumber \\
 &\qquad & E_\Delta = \frac{16}{9} \frac{\hbar^2\pi^2}{2 \mu L^2} ~~~~ m = 1,2,3,...\\
 &\qquad & n = 1,2,3,... (m \geq 2n).\nonumber
\end{eqnarray}
All the eigenvalues are doubly degenerate except ($m=2n$) which is singly degenerate. Taking these in consideration the level density is given by
\begin{eqnarray}\label{eq:trige}
g(E) &=& \frac{\pi}{3\sqrt{3} E_\Delta} \sum_{M_1,M_2 = -\infty}^{\infty} J_0(kL_{M_1,M_2})-\frac{1}{2\sqrt{E E_\Delta}} \sum_{M=-\infty}^{\infty} \cos (k L_{M_1,M_2}) +\frac{1}{3} \delta(E) \nonumber \\
 &\qquad & \mbox{where  }  L_{M_1,M_2}= \sqrt{3(M_1^2+M_2^2+M_1 M_2)}L.
\end{eqnarray}
The analogue of (\ref{eq:5}) and (\ref{eq:6}) for equilateral triangular billiards are 

\begin{eqnarray}\label{eq:triNN0}
N_n(E) - N_0(E) &=& \frac{\pi}{3\sqrt{3}E_\Delta} \sum_{(M_1=0,M_2=1)}^{\infty} \frac{8\sqrt{E}}{L_{0,M_2}}J_1[\sqrt{E}L_{0,M_2}] \nonumber \\&~&+\frac{\pi}{3\sqrt{3}E_\Delta}\sum_{M_1,M_2=1}^\infty \frac{4\sqrt{E}}{L_{M_1,M_2}}J_1[\sqrt{E}L_{M_1,M_2}] \nonumber \\
&~&+\frac{\pi}{3\sqrt{3}E_\Delta}\sum_{M_1=1}^\infty \sum_{M_1=-\infty}^{-1} \frac{4\sqrt{E}}{L_{M_1,M_2}}J_1[\sqrt{E}L_{M_1,M_2}] \nonumber \\&~&-\frac{2}{E_\Delta }\sum_{M=1}^{\infty} \frac{1}{L_M} \sin (\sqrt{E} L_M )
\end{eqnarray}
\begin{eqnarray}\label{eq:trin}
n(E\leq E_0) &=& \frac{\pi E_0}{3\sqrt{3} E_\Delta}-\sqrt{\frac{E_0}{E_\Delta}} + \frac{1}{3}.
\end{eqnarray}

In our calculation of stochasticity parameter for this system $E_\Delta = \pi^2$. A comparison similar to rectangular billiard between the stochasticity parameter calculated exactly using $5000$ energy levels of equilateral triangular billiard and using trace formula is presented in Fig.\ref{fig:sp_tri}.

\noindent
{\it Zeros of Riemann zeta function}

Riemann zeta function is a very important special function \cite{edwards} with many interesting connections with number theory. It is defined in terms of a Dirichlet series or a product form as follows:
\begin{eqnarray}\label{eq:8}
\zeta (s) &=& \sum_{n=1}^{\infty } \frac{1}{n^s} \nonumber \\
&=& \prod_{\mbox{primes,~}p} \left( 1 - \frac{1}{p^{\sigma + it}} \right)^{-1}.
\end{eqnarray}
According to the Riemann hypothesis, all the nontrivial zeros lie on the critical line, given by $s_n = \frac{1}{2}+it_n$. The product formula is convergent only for $\sigma > 1$, but we employ it to write an analogue of Gutzwiller trace formula  \cite{gutzwiller}. Taking logarithm of the product form above and using $\log (1-x) = - \sum_{k=1}^{\infty }x^k/k$, 
\begin{equation}\label{eq:9}
\log \zeta (s) = \sum_{p} \sum_{k} \frac{\exp (-ikt\log p)}{kp^{k\sigma }}.
\end{equation}
Taking imaginary part of both sides, and differentiating with respect to $t$, we obtain for fixed $\sigma > 1$:
\begin{equation}\label{eq:10}
d_{osc}^{\sigma } = - \frac{1}{\pi }\sum_{p} \sum_{k} \frac{\log p}{p^{k\sigma }}\cos (kt \log p).
\end{equation}
The oscillating part of the cumulative density will be
\begin{equation}\label{eq:11}
N_{osc}^{\sigma } = - \frac{1}{\pi }\sum_{p} \sum_{k} \frac{1}{kp^{k\sigma }}\sin (kt \log p).
\end{equation}
For the denominator of (\ref{eq:2}), 
\begin{eqnarray}\label{eq:12}
n &=& \mbox{Int}\left[ \frac{t}{2\pi }\log \frac{t}{2\pi }-\frac{t}{2 \pi}\right]. 
\end{eqnarray}
Finally, for the zeros of Riemann zeta functions on the critical line, 
\begin{eqnarray}\label{eq:13}
\lambda &=& \sup_{t}\frac{|N_{osc}^{\sigma }|}{\sqrt{n}} \nonumber \\
&=& -\sup_{t} \sqrt{\frac{2}{\pi}} \bigg| \sum_{p} \sum_{k=1}^{\infty } \frac{\sin [(k\log p)t]}{k(p)^{k\sigma }\sqrt{t\log (t/2\pi )-t}} \bigg|.
\end{eqnarray} 
This expression should be doomed as its convergence is in the domain where there are no nontrivial zeros. However, as noted in \cite{brack1}, trace formula contains information about the zeros on the critical line, with the shortest orbits (corresponding to the smallest primes) contributing the most \cite{arul}. We have compared the values of stochasticity parameter using numerical values of the zeros (www.dtc.umn.edu/$\sim $odlyzko/zeta\_tables/) and the expression (\ref{eq:13}) in Fig.\ref{fig:sp_zeta}.  For calculation we have utilized first $5000$ only. This is done to compare the results from various systems for same sequence length.


\begin{thebibliography}{99}
\bibitem{footnote1} For completeness, it is good to write the definition of the stochasticity parameter, $\lambda _n$ as give by Arnold for real numbers, $x_1 \leq x_2 \leq ... \leq x_n$. Let the number of elements $x_i$ which are $\leq X$ be given by an empirical counting function, $C_n(X)$. Let us assume that there is a theoretical counting function, $C_0(X) = n.$ (probability that an event $x \leq X$). The parameter is defined as
$\lambda _n = \sup_{X} (C_n(X) - C_0(X))/\sqrt{n}$. The value $\lambda _n$ itself is a random variable and the distribution function of $\lambda _n$, $n \to \infty $ converges to the universal function, the Kolmogorov distribution, Eq. \ref{kolmo}.
\end{thebibliography}
\end{document}